\documentstyle[aps,prd,amssymb]{revtex}

\def\np{({\bf n}\cdot{\bf p})}
\def\pp{{\bf p}^2}
\def\ppp{({\bf p}^2)}

\def\dpdp{\dot{\bf p}^2}
\def\pdp{({\bf p}\cdot\dot{\bf p})}
\def\drdr{\dot{\bf r}^2}
\def\dpdr{(\dot{\bf p}\cdot\dot{\bf r})}
\def\pdr{({\bf p}\cdot\dot{\bf r})}
\def\ndp{({\bf n}\cdot\dot{\bf p})}
\def\ndr{({\bf n}\cdot\dot{\bf r})}

\def\ppa{({\bf p}_a)^2}
\def\nabpa{({\bf n}_{ab}\cdot{\bf p}_a)}

\def\omk{\omega_{\text{kinetic}}}
\def\oms{\omega_{\text{static}}}

\begin{document}

\title{The binary black-hole dynamics
at the third post-Newtonian order in the orbital motion}

\author{Piotr Jaranowski}
\address{Institute of Theoretical Physics,
University of Bia{\l}ystok,
Lipowa 41, 15-424 Bia{\l}ystok, Poland}

\author{Gerhard Sch\"afer}
\address{Theoretisch-Physikalisches Institut,
Friedrich-Schiller-Universit\"at,
Max-Wien-Pl. 1, 07743 Jena, Germany}

\maketitle

\begin{abstract}
The orbital dynamics of the binary point-mass systems is ambiguous at the third
post-Newtonian order of approximation.  The static ambiguity is known to be
related to the difference between the Brill-Lindquist solution and the
Misner-Lindquist solution of the time-symmetric conformally flat initial value
problem for binary black holes.  The kinetic ambiguity is noticed to violate in
general the standard relation between the center-of-mass velocity and the total
linear momentum as demanded by global Lorentz invariance.
\end{abstract}

\pacs{04.70.Bw, 04.20.Ex, 04.20.Fy}

\section{Introduction}

The motion of binary systems is one of the most important problems in 
general
relativity, particularly in view of the future detection of gravitational 
waves
from such systems.  The simplest two-body problem is the one where the
components of the system are objects which are as spherically symmetric 
and
point-like as possible.  In general relativity, those objects are black 
holes.
In spite of the extension of black holes, in binary orbit approximate
calculations of post-Newtonian (PN) type (which use expansions in powers 
of
$1/c$, where $c$ denotes the velocity of light), the ansatz of Dirac 
delta
functions to describe point-like sources turned out to be very successful 
up to
the 2.5PN approximation \cite{DD81,D82}, and at the 3.5PN order of 
approximation
too \cite{js97}.  At the 3PN order, i.e.\ at the order $(1/c^2)^3$, only 
partial
success was achievable \cite{js98a}.  Two terms in the binary point-mass
calculations came out ambiguously:  a kinetic one, depending on the 
bodies'
linear momenta squared \cite{js98a}, and a static one, not depending on 
the
momenta \cite{js99}.  The authors were able to show that the static 
ambiguity is
intimately related to the fact that different exact solutions 
(Brill-Lindquist
and Misner-Lindquist) of the time-symmetric and conformally flat initial 
value
problem for two black holes do exist, which at the time reveal the 
emergence of 
black holes in binary point-mass calculations at the 3PN order 
\cite{js99}. In 
this paper we confront the kinetic ambiguity with the center-of-mass 
motion and 
claim that it can be fixed through the standard relation between the 
center-of-mass 
velocity and the total linear momentum which results from global Lorentz 
invariance.

We employ the following notation:  ${\bf x}_a=\left(x_a^i\right)$ 
($a=1,2$; 
$i=1,2,3$) denotes the position of the $a$th point mass in the 
3-dimensional 
Euclidean space endowed with a standard Euclidean scalar product (denoted 
by a 
dot). We also define
${\bf r}_{12}\equiv{\bf x}_1-{\bf x}_2$,
$r_{12}\equiv|{\bf r}_{12}|$,
${\bf n}_{12}\equiv{\bf r}_{12}/r_{12}$;
$|\cdot|$ stands here for the Euclidean length of a vector.  The linear 
momentum 
of the $a$th body is denoted by ${\bf p}_a=\left(p_{ai}\right)$.

\section{The generalized conservative Hamiltonian to 3PN order}

The post-Newtonian approximation order 2.5PN (as well 3.5PN) is of purely
dissipative character.  Subtraction of the 2.5PN terms from the 3PN 
Hamiltonian
results in the conservative 3PN Hamiltonian.  This Hamiltonian is 
primarily of
higher order as it depends on time derivatives of the positions and 
momenta of
the bodies \cite{js98a}.

The conservative 3PN higher-order Hamiltonian for two-body point-mass 
systems 
was calculated in Ref.\ \cite{js98a}. We present here the explicit form 
of this 
Hamiltonian in the center-of-mass reference frame (where
${\bf p}_1+{\bf p}_2=0$). It is convenient to introduce the following 
reduced 
variables
\begin{equation}
\label{h1}
{\bf r} \equiv \frac{{\bf x}_1-{\bf x}_2}{GM},\quad
{\bf p} \equiv \frac{{\bf p}_1}{\mu} = -\frac{{\bf p}_2}{\mu},\quad
\widehat{t} \equiv \frac{t}{GM},\quad
\widehat{H}^{\rm NR} \equiv \frac{H^{\rm NR}}{\mu},
\end{equation}
where
\begin{equation}
\label{h2}
\mu \equiv \frac{m_1m_2}{M},\quad M \equiv m_1 + m_2,\quad
\nu \equiv \frac{\mu}{M} = \frac{m_1m_2}{(m_1+m_2)^2}.
\end{equation}
In Eq.\ (\ref{h1}) the superscript NR denotes a `non-relativistic' 
Hamiltonian, 
i.e.\ the Hamiltonian without the rest-mass contribution $Mc^2$. We also 
introduce the reduced-time derivatives:
\begin{equation}
\dot{\bf r} \equiv \frac{d{\bf r}}{d\widehat{t}}
= \frac{d({\bf x}_1-{\bf x}_2)}{dt},\quad
\dot{\bf p} \equiv \frac{d{\bf p}}{d\widehat{t}}
= \frac{G}{\nu}\frac{d{\bf p}_1}{dt} = -\frac{G}{\nu}\frac{d{\bf 
p}_2}{dt}.
\end{equation}
The reduced two-point-mass conservative Hamiltonian up to the 3PN order 
reads
\begin{eqnarray}
\label{hnr}
\widehat{H}^{\text{NR}}\left({\bf r},{\bf p},\dot{\bf r},\dot{\bf 
p}\right)
&=& \widehat{H}_{\rm N}\left({\bf r},{\bf p}\right)
+ \frac{1}{c^2} \widehat{H}_{\rm 1PN}\left({\bf r},{\bf p}\right)
\nonumber\\[2ex]&&
+ \frac{1}{c^4} \widehat{H}_{\rm 2PN}\left({\bf r},{\bf p}\right)
+ \frac{1}{c^6}
\widehat{H}_{\rm 3PN}\left({\bf r},{\bf p},\dot{\bf r},\dot{\bf 
p}\right),
\end{eqnarray}
where (here $r\equiv|{\bf r}|$ and ${\bf n}\equiv{\bf r}/r$)
\begin{equation}
\widehat{H}_{\rm N}\left({\bf r},{\bf p}\right) = 
\frac{\pp}{2}-\frac{1}{r},
\end{equation}
\begin{equation}
\widehat{H}_{\rm 1PN}\left({\bf r},{\bf p}\right)
= \frac{1}{8}(3\nu-1)\ppp^2
- \frac{1}{2}\left[(3+\nu)\pp+\nu\np^2\right]\frac{1}{r} + 
\frac{1}{2r^2},
\end{equation}
\begin{eqnarray}
\widehat{H}_{\rm 2PN}\left({\bf r},{\bf p}\right) &=& 
\frac{1}{16}\left(1-5\nu+5\nu^2\right)\ppp^3
\nonumber\\[2ex]&&
+ \frac{1}{8}
\left[\left(5-20\nu-3\nu^2\right)\ppp^2-2\nu^2\np^2\pp-3\nu^2\np^4\right]
\frac{1}{r}
\nonumber\\[2ex]&&
+ \frac{1}{2}\left[(5+8\nu)\pp+3\nu\np^2\right]\frac{1}{r^2}
- \frac{1}{4}(1+3\nu)\frac{1}{r^3},
\end{eqnarray}
\begin{eqnarray}
\label{H3PN}
\widehat{H}_{\rm 3PN}\left({\bf r},{\bf p},\dot{\bf r},\dot{\bf p}\right)
&=& \frac{1}{128}\left(-5+35\nu-70\nu^2+35\nu^3\right)\ppp^4
\nonumber\\[2ex]&&
+ \frac{1}{16} \Big\{ \left(-7+42\nu-53\nu^2-6\nu^3\right)\ppp^3
\nonumber\\[2ex]&&
+ (1-2\nu)\nu^2 \left[2\np^2\ppp^2 + 3\np^4\pp\right] \Big\}\frac{1}{r}
\nonumber\\[2ex]&&
+\frac{1}{48} \Big[ 3\left(-27+140\nu+96\nu^2\right)\ppp^2
\nonumber\\[2ex]&&
+ 6(8+25\nu)\nu\np^2\pp - (35-267\nu)\nu\np^4 \Big]\frac{1}{r^2}
\nonumber\\[2ex]&&
+\frac{1}{1536} \Big\{
\left[-4800-3(8944-315\pi^2)\nu-7808\nu^2\right]\pp
\nonumber\\[2ex]&&
+ 9\left(2672-315\pi^2+448\nu\right)\nu\np^2 \Big\}\frac{1}{r^3}
\nonumber\\[2ex]&&
+\frac{1}{96} \Big[12+(872-63\pi^2)\nu\Big]\frac{1}{r^4}
\nonumber\\[2ex]&&
+\widehat{D}\left({\bf r},{\bf p},\dot{\bf r},\dot{\bf p}\right)
+\widehat{\Omega}\left({\bf r},{\bf p}\right).
\end{eqnarray}

In Eq.\ (\ref{H3PN})
$\widehat{D}\left({\bf r},{\bf p},\dot{\bf r},\dot{\bf p}\right)$ denotes 
that 
part of the 3PN Hamiltonian $\widehat{H}_{\rm 3PN}$ which depends on the 
time 
derivatives $\dot{\bf r}$ and $\dot{\bf p}$. Its explicit form reads
\begin{equation}
\widehat{D} = \widehat{D}_1\,r + \widehat{D}_0 + 
\widehat{D}_{-1}\,\frac{1}{r}
+ \widehat{D}_{-2}\,\frac{1}{r^2} + \widehat{D}_{-3}\,\frac{1}{r^3},
\end{equation}
where
\begin{eqnarray}
\widehat{D}_1 &=& \frac{1}{12}\nu^2 \Big[
4\np\ndp\pdp - 5\ndp^2\pp
\nonumber\\[2ex]&&
- 5\np^2\dpdp - \np^2\ndp^2 + 13\pp\dpdp + 2\pdp^2 \Big],
\end{eqnarray}
\begin{eqnarray}
\widehat{D}_0 &=& \frac{1}{8}\nu^2 \Big\{
\ndr\pdp \big[5\pp+\np^2\big]
\nonumber\\[2ex]&&
- \np \big[\pp\dpdr+2\pdp\pdr\big] - \frac{1}{3}\np^3\dpdr
\nonumber\\[2ex]&&
+ \ndp \big[\pp+\np^2\big] \big[\np\ndr-\pdr\big]
\Big\},
\end{eqnarray}
\begin{eqnarray}
\widehat{D}_{-1} &=& \frac{1}{24}\nu \Big\{
2(17-10\nu)\np\ndp\ndr - (15+22\nu)\np^2\ndp
\nonumber\\[2ex]&&
- (51-8\nu)\ndp\pp - 2(6-5\nu)\np\pdp
\nonumber\\[2ex]&&
-2(1-2\nu)\ndr\pdp - 2(7-2\nu)\big[\np\dpdr+\ndp\pdr\big] \Big\}
\nonumber\\[2ex]&&
+ \frac{1}{16}\nu^3 \Big[ 8\np\ndr\pp\pdr + 8\np^3\ndr\pdr
\nonumber\\[2ex]&&
+2\np^2\pp\drdr + 5\ppp^2\drdr + \np^4\drdr - 5\ndr^2\ppp^2
\nonumber\\[2ex]&&
-6\np^2\ndr^2\pp - 5\np^4\ndr^2
\nonumber\\[2ex]&&
-4\np^2\dpdr^2 - 4\pp\dpdr^2 \Big],
\end{eqnarray}
\begin{eqnarray}
\widehat{D}_{-2} &=& \frac{1}{48}\nu \Big[
5(5-7\nu)\np^3\ndr + 10(3-5\nu)\np^2\ndr^2
\nonumber\\[2ex]&&
+ 3(17-35\nu)\np\ndr\pp - 28(3-8\nu)\np\ndr\pdr
\nonumber\\[2ex]&&
- 15(2-3\nu)\np^2\pdr + 2(24-77\nu)\np^2\drdr
\nonumber\\[2ex]&&
+ 2(9-29\nu)\ndr^2\pp - 3(4-9\nu)\pp\pdr
\nonumber\\[2ex]&&
- 2(12-37\nu)\pp\drdr + 4(6-17\nu)\pdr^2
\Big],
\end{eqnarray}
\begin{eqnarray}
\widehat{D}_{-3} &=& \frac{1}{1536}\nu \Big\{
3\big[927\pi^2-10832+36\left(48-5\pi^2\right)\nu\big]\np\ndr
\nonumber\\[2ex]&&
- 6\big[3(16+\pi^2)-2(45\pi^2-464)\nu\big]\ndr^2
\nonumber\\[2ex]&&
+ \big[11600-927\pi^2+20(9\pi^2-80)\nu\big]\pdr
\nonumber\\[2ex]&&
+ 2\big[176+3\pi^2+6(176-15\pi^2)\nu\big]\drdr
\Big\}.
\end{eqnarray}

The last term in Eq.\ (\ref{H3PN}),
$\widehat{\Omega}\left({\bf r},{\bf p}\right)$, 
contains the parameters which parametrize the ambiguities of the 3PN 
Hamiltonian. Its explicit form is
\begin{equation}
\label{Omega}
\widehat{\Omega}\left({\bf r},{\bf p}\right)
= \omk \big[\pp-3\np^2\big]\frac{\nu^2}{r^3} + \oms \frac{\nu}{r^4},
\end{equation}
where $\omk$ parametrizes the kinetic ambiguity, and the static ambiguity 
is 
pa\-ra\-me\-trized by $\oms$.

The 3PN higher-order Hamiltonian
$\widehat{H}_{\rm 3PN}\left({\bf r},{\bf p},\dot{\bf r},\dot{\bf 
p}\right)$, 
Eq.\ (\ref{H3PN}), can be reduced to a usual Hamiltonian depending only 
on
${\bf r}$ and ${\bf p}$. Details of the reduction can be found in Sec.\ 
II of 
Ref.\ \cite{djs99}.

\section{The ambiguous part of the 3PN Hamiltonian}

It was found in \cite{js98a}, \cite{js99} that the 
ambiguous part $\Omega$ of the 3PN Hamiltonian, Eq.\ (\ref{Omega}), 
takes, 
in a non-center-of-mass reference frame and in non-reduced variables, the 
form 
($G$, the Newtonian gravitational constant, is put equal to one)
\begin{eqnarray}
\label{Omeganr}
\Omega({\bf{x}}_1,{\bf{x}}_2,{\bf{p}}_1,{\bf{p}}_2)
&=& \frac{1}{c^6} \sum_a\sum_{b\ne a} \left\{
\oms \frac{m_a^3 m_b^2}{r^4_{ab}}
\right.\nonumber\\[2ex]&&\left.
+ \frac{1}{2}\omk \frac{m_a\,m_b}{r_{ab}^3} \left[\ppa-3\nabpa^2\right]
\right\}.
\end{eqnarray}

In the following we shall show how center-of-mass considerations can 
impose 
restrictions onto the ``kinetic'' ambiguous term.
To make contact with the intended center-of-mass motion calculations we  
transform (\ref{Omeganr}) to the Lagrangean level. On this level, 
$\Omega$ 
appears as $-\Omega$ with 
\begin{eqnarray}
\label{Omegala}
\Omega({\bf{x}}_1,{\bf{x}}_2,{\bf{v}}_1,{\bf{v}}_2)
&=& \frac{1}{c^6} \sum_a\sum_{b\ne a} \left\{
\oms \frac{m_a^3 m_b^2}{r^4_{ab}}
\right.\nonumber\\[2ex]&&\left.
+ \frac{1}{2}\omk \frac{m^3_a\,m_b}{r_{ab}^3} 
\left[({\bf{v}}_a)^2-3({\bf{n}}_{ab}{\bf{v}}_a)^2\right]
\right\},
\end{eqnarray}
where the bodies' velocities are given by ${\bf{v}}_a = {\bf{p}}_a/m_a$.

In view of the relation between the total linear momentum ${\bf{P}}$ and 
the 
center-of-mass velocity ${\bf{V}} = \dot{\bf{X}}$ (${\bf{X}}$ denotes the 
center-of-mass coordinate) demanded by global Lorentz invariance, 
\cite{dd81a},
\begin{equation}
\label{CM}
{\bf{P}} = \frac{H}{c^2} \frac{d{\bf{X}}}{dt}, \quad {\bf{P}} = 
{\bf{p}}_1 + 
{\bf{p}}_2,
\end{equation}
where $H$ is the conserved total energy, $H = Mc^2 + H^{\text{NR}}$, the 
following expression (which is one of the contributions to 
${\bf{P}} = \partial L / \partial {\bf{v}}_1 + \partial L / \partial 
{\bf{v}}_2$)
\begin{eqnarray}
\label{Omegall}
{\bf{D}} \equiv - \frac{\partial \Omega}{\partial {\bf{v}}_1}
 - \frac{\partial\Omega}{\partial {\bf{v}}_2}
= - \omk \frac{1}{c^6} \sum_a\sum_{b\ne a} 
\frac{m^3_a\,m_b}{r_{ab}^3} 
\left[{\bf{v}}_a-3({\bf{n}}_{ab}{\bf{v}}_a){\bf{n}}_{ab}\right]
\end{eqnarray}
has either to be a total time derivative (in which case no restrictions 
are 
imposed on $\Omega$) 
or to be fixed by other terms in Eq.\ (\ref{CM}) (note
that no static ambiguity is involved here).

The simplest way to show that ${\bf{D}}$ is not a total time derivative 
is to 
write it in the form 
\begin{equation}
\label{td}
D^i = \omk m_1m_2 [(m_1^2 + m_2^2) V^j + \frac{m_1m_2(m_1-m_2)}{M} v^j] 
\partial_{1j} \partial_{1i} \frac{1}{r_{12}}, 
\end{equation}
($v^i$ denotes the relative velocity, $v^i = v^i_1-v^i_2$, and here, 
$MV^i=m_1v^i_1 +m_2v^i_2$ holds)  
and to average $D^i$ over a circular orbit of the relative motion simply 
assuming that the center-of-mass 
motion is orthogonal to the relative motion. The result is (notice that 
the relative motion part in the 
Eq. (\ref{td}) is a total time derivative)  
\begin{equation}
<{\bf{D}}> =  - \omk\frac{m_1m_2}{r^3_{12}}(m_1^2+m_2^2){\bf{V}}.
\end{equation}
In the center-of-mass frame $D^i$ is always a total time derivative. This 
fits 
with Ref.\ \cite{js98b} where the kinetic ambiguity, in the 
center-of-mass 
frame, was shown to be influenced by coordinate transformations which 
additionally only influence, unimportant in this context, the static 
potential 
at 3PN order.

\section{Conclusions}

In this paper we have shown that the kinetic ambiguity obtained in
binary point-mass calculations at the 3PN order should be getting fixed 
by
center-of-mass motion considerations. The static ambiguity was known to 
be 
fixable by referring to e.g., the Brill-Lindquist ($\oms =0$) or 
the Misner-Lindquist solution ($\oms = -1/8$). The energy content of the 
latter
solution is influenced by the topology of the involved non-simply 
connected 
3-space.

%%%%%%%%%%%%%%%%%%%%%%%%%%%%%%%%%%%%%%%%%%%%%%%%%%%%%%%%%%%%%%%%%%%%%%%%%
%%
%%%%%%%%%%%%%%%%%%%%%%% Acknowledgements 
%%%%%%%%%%%%%%%%%%%%%%%%%%%%%%%%%%
%%%%%%%%%%%%%%%%%%%%%%%%%%%%%%%%%%%%%%%%%%%%%%%%%%%%%%%%%%%%%%%%%%%%%%%%%
%%
\vspace*{0.25cm}\baselineskip=10pt{\small\noindent
G.S. wishes to acknowledge useful discussions with Luc Blanchet about 
Lorentz invariance. The authors also thank Thibault Damour for
constructive remarks. This work was supported in part by the 
Max-Planck-Gesellschaft Grant No.\ 
02160-361-TG74 (G.S.) and the KBN Grant No.\ 2 P03B 094 17 (P.J.).}
%%%%%%%%%%%%%%%%%%%%%%%%%% Bibliography 
%%%%%%%%%%%%%%%%%%%%%%%%%%%%%%%%%%%
%%%%%%%%%%%%%%%%%%%%%%%%%%%%%%%%%%%%%%%%%%%%%%%%%%%%%%%%%%%%%%%%%%%%%%%%%
%%

\end{document}